# FOIL SCATTERING MODEL FOR FERMILAB BOOSTER*

J. Eldred, C. M. Bhat, S. Chaurize, V. Lebedev, S. Nagaitsev, K. Seiya, C. Y. Tan, R. J. Tesarek
Fermilab, Batavia, IL 60510, USA

*Abstract*

At the Fermilab Booster, and many other proton facilities, an intense proton beam is accumulated by multi-turn injection of an H- beam through a stripping foil. The circulating beam scatters off the injection foil and large-angle Coulomb scattering leads to uncontrolled losses concentrated in the first betatron period. We measure the foil scattering loss rate in the Fermilab Booster as a function of LINAC current, number of injection turns, and time on the injection foil. We find that current Booster operation has ~1% foil scattering loss and we make projections for the Proton Improvement Plan II (PIP-II) injector upgrade. Here we present the results from our recent beam measurements and a foil scattering model analyses.

## INTRODUCTION

The Fermilab Booster's near-term improvement plan includes a 20% increase in intensity from $4.5\times10^{12}$ protons to $5.4\times10^{12}$ protons per Booster batch (ppBc) which would be enabled by a commensurate reduction in beam loss radio-activation. The quantification and accurate identification of Booster loss mechanisms forms an integral part of the loss reduction strategy. In this paper we examine the subset of Booster injection losses which are directly attributable to scattering of the incoming and circulating beam off the injection foil.

Large-angle Coulomb scattering from the injection foil is an operational reliability concern for the Fermilab Booster. The Booster has about a 0.57 π phase-advance per period and consequently particles with extreme divergence at the injection foil will primarily be lost in the downstream long straight section. That straight section includes extraction kickers which have had shortened service lifetimes due to severe radiation damage.

The study will also inform a later PIP-II intensity upgrade [1], in which the Booster will use a new 800 MeV injection region with a transversely painted foil-injection process. The PIP-II injection region design is actively being optimized to minimize and control the localized injection losses. Not only at Fermilab, at other proton facilities – such as SNS, J-PARC, and Los Alamos, including the current injection region in the Fermilab Booster – where the highest radio-activation levels are observed at the injection region [2–4]. We find foil-scattering injection loss to be a subject of broad interest.

## SCATTERING MEASUREMENTS

Many scintillator and photo-multiplier tube (PMT) detectors were installed early 2015 around the collimation region of the Booster to identify and monitor beam loss mechanisms [5]. The response time of the PMT detectors is a few nanoseconds, allowing fast-loss mechanisms such as injection foil scattering to be visualized in detail. After three years of operation, several PMT detectors were non-operational due to radiation damage and are in the process of being replaced. The PMT detector attached to the "5-1 downstream dipole magnet" is operational however and provides a clear sample of the injection foil scattering loss.

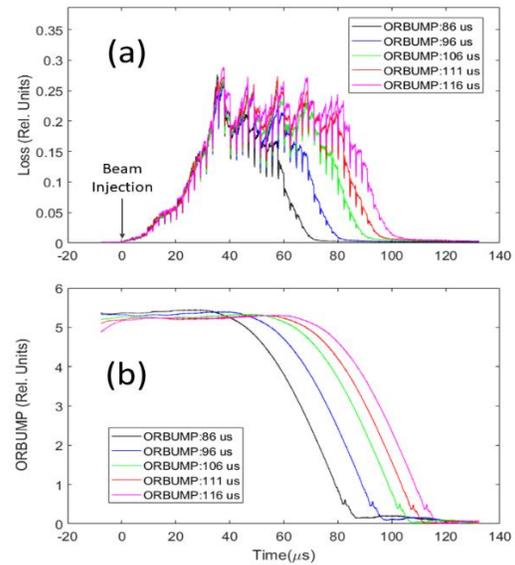

Figure 1: (a) PMT signal for first 140 μs at 14-Booster turn beam injection with LINAC current of 24 mA. The black curve with legend "ORBUMP:86 μs" corresponds to beam on foil for ≈70 μs. (b) ORBUMP data.

The PMT foil-scattering signal was studied under varying injection conditions. The PMT foil-scattering signal was measured with the number of revolutions injected into the Booster at 10-turns and 14-turns ($3.3\times10^{12}$ ppBc and $4.7\times10^{12}$ ppBc at nominal current of 25mA). To obtain data at reduced current of 10 mA and 1 mA, the LINAC beam is collimated downstream of the RFQ. The PMT foil-scattering signal was also measured by varying the duration that the circulating beam is on the foil from 2.2$BT$+40 μs (nominal) to 2.2$BT$+70 μs. The quantity $BT$ is number of turns during multi-turn injection. For each case, the PMT signal for 64 consecutive accelerator cycles are averaged to minimize the impact of cycle-to-cycle variation.

Fig. 1(a) shows averaged PMT signals for nominal beam current of 14-turn injection. The losses are associated with foil scattering, rising when the beam is injected and falling as the circulating beam moves off the injection foil. The rise of the scattering signal is not simply linear, which is consistent with ripple in the Booster corrector magnet power supplies. The sudden decreases in the foil-scattering

---

signal at every 2.2 μs correspond to a "notch" in the beam pulse made in downstream of RFQ using a laser notcher system [6]. Figure 1(b) shows the corresponding ORBUMP setting. Longer ORBUMP implies a greater number of circulating beam hits on the foil.

## FOIL-SCATTERING LOSS ESTIMATE

The PMT voltage determines the overall gain of the detector, which was set to maximize the signal-to-noise without saturating the signal. Data were taken at LINAC beam current of 24-mA, 10 mA, and 1 mA with PMT gain of 1540 V, 1740 V, and 2000 V, respectively. For each current, one data set was taken at two voltages for calibration and all data were normalized to 1740 V data. After gain-calibration and baseline subtraction, the signal is integrated to determine the foil-scattering for each injection condition. The PMT samples an unknown fraction of the total foil-scattering losses, and therefore indicates the relative changes in scattering under the injection conditions.

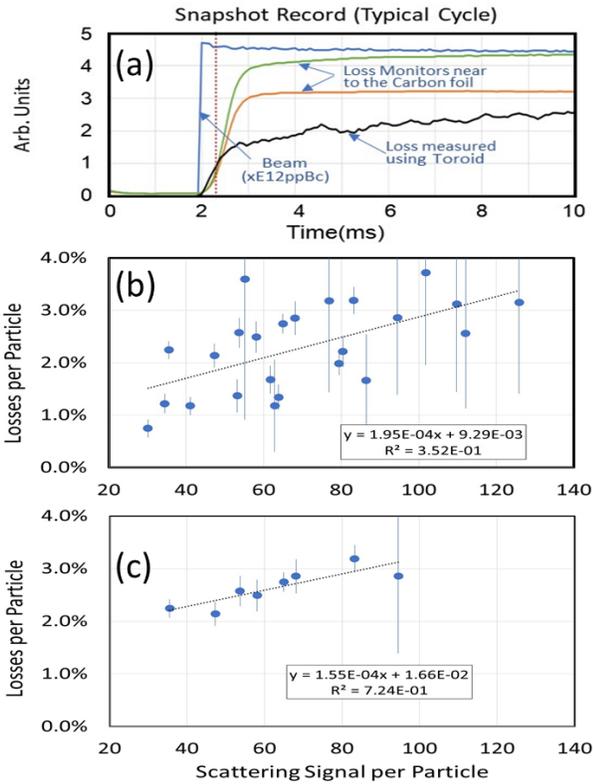

Figure 2: (a) Operational beam and loss. Loss measured by toroid, calculated as a decline in peak charge, with 10x magnification and 0.3 ms averaging window. The 320 μs measurement time indicated by vertical dotted line. (b) Correlation between the measured loss per particle with toroid and the average signal per particle measured using PMT with linear fit. Datapoints across all injection conditions. (c) Datapoints for 24 mA LINAC current only. The error bars are due to data sample size.

Booster beam and loss monitor data on a typical beam cycle is shown in Fig. 2(a). To estimate the total losses due to foil scattering, we can correlate that signal to the absolute beam loss after injection measured using a toroid. We calculate the absolute losses from the decline in measured beam intensity 320 μs after injection. When losses increase as a function of the duration that the circulating beam is left on the injection foil, this is a strong indicator that those losses can be attributed to foil scattering. To combine the data across different LINAC currents and injected turns however, the measured beam loss and the scattering signal must both be normalized by beam intensity.

The correlation between the toroid-measured loss per particle and the scattering signal per particle across all injection conditions are shown in Figs. 2(b). Figure 2(c) shows the correlation only for the eight data-points with the nominal settings of 24 mA. For the first correlation plots the $R^2$ coefficient of determination is only 0.35, but the value is 0.74 for the second correlation plots. The error bars in these plots are primarily coming cycle-to-cycle variation and small number of a samples.

The vertical-intercept of the correlation plots represents an estimate of the injection losses from unrelated mechanisms. The losses proportional to the scattering signal are an estimate of the losses attributable to foil-scattering mechanisms. We estimate that for nominal injection conditions ($14BT@24$ mA) to have 1.1% foil scattering losses from Fig. 2(a) and ~0.83% foil-scattering losses from Fig. 2(b). We take the two estimates as an indirect measure of systematic bias, obtain an average estimate of ~1% foil-scattering loss for nominal operation. At 400 MeV this beam loss amounts to 3 J out of the ring-wide loss budget of 33 J.

## FOIL-SCATTERING LOSS MODEL

Stripping of 800 MeV H$^-$ beam through carbon foil was studied in detail by M. S. Gulley, et. al., [7] which addresses final product H$^-$ without any interaction, H$^0$, excited states of H$^0$ and H$^+$. In the context of our recent study at 400 MeV to characterize the beam loss mechanism during multi-turn beam injection of H$^-$ stripping by a carbon foil and planned PIP-II at 800 MeV injection, we revisited many of the issues, like beam loss through nuclear scattering, emittance growth through multiple Coulomb scattering and beam loss through large angle Coulomb scattering. Figure 3(a) presents estimated probability for various processes. In the current operational scenario, we expect that H$^-$ → H$^+$ stripping efficiency is ~99.89% and that during PIP-II we plan to achieve stripping efficiency of 99.99% by using a thicker carbon foil of ~500 μgm/cm$^2$. However, currently the circulating H$^+$ beam under goes a minimum of 19 hits with the stripping foil, causing unwanted beam losses. Using nuclear scattering model [8] correcting it for multi-turn beam injection, we found the beam loss due to nuclear scattering is ~0.02%/proton. On the other hand, that during the PIP-II era number of hits will be ~6 and expected loss will be ~0.002%/proton. The loss arising from large angle Coulomb scattering in all the above cases are non-negligible and hence, addressed here in detail.

For an injected beam with zero emittance on a thin foil, the probability $P$ for losses of proton due to large angle

Coulomb scattering during multi-turn injection can be expressed in terms of foil hits $n_{hit}$ (number of hits beam makes after the completion of the injection process), foil thickness $x$, and acceptance angles $\theta_{xl}$ and $\theta_{yl}$ at the injection location [4]:

$$P = \Upsilon x \left\{ n_{hit} + \frac{BT+1}{2} \right\} \quad (1).$$

where,

$$\Upsilon = \rho \left( \frac{2 Z m_e r_e}{\gamma m_p \beta^2} \right)^2 \left[ \frac{1}{\theta_{xl} \theta_{yl}} + \frac{1}{\theta_{xl}^2} \operatorname{atan}\left( \frac{\theta_{yl}}{\theta_{xl}} \right) + \frac{1}{\theta_{yl}^2} \operatorname{atan}\left( \frac{\theta_{xl}}{\theta_{yl}} \right) \right] \quad (2).$$

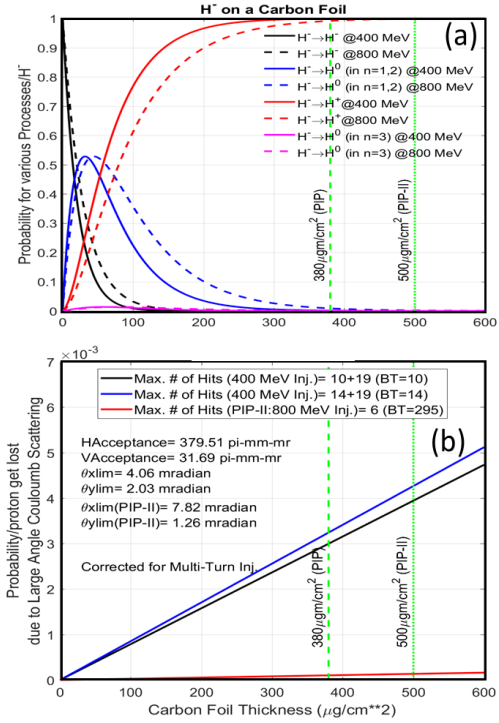

Figure 3: (a) Probability for various end products due to one pass interaction of 400 MeV and 800 MeV H⁻ beam on a carbon foil as a function of foil thickness. (b) Estimated probability for proton that get lost due to large angle Coulomb scattering as a function of carbon foil thickness.

The quantities, $x$, $\rho$, $Z$, $m_e$, $r_e$ and $m_p$ are foil thickness in μg/cm², number of foil atoms/gm, electron mass, classical radius of the electron, mass of the circulating proton, respectively. $\gamma$ and $\beta$ are relativistic quantities. The Eq. (2) assumes a rectangular aperture. For a foil-injection offset from the circulating beam orbit, the acceptance angle $\theta_y$ is constrained by:

$$2\beta_y J_{y,max} = y_i^2 + (\alpha_y y + \beta_y \theta_y)^2 \quad (3).$$

where $J_{y,max}$ is the maximum vertical action accepted by the ring. $\beta_y$ and $\alpha_y$ are the TWISS functions at the injection foil, and $y_i$ is the vertical orbit offset. A similar expression applies for the horizontal plane. For a beam of non-negligible emittance Eq. (1) becomes cumbersome and a more general expressions can be derived. The physics treatment becomes analogous to the case of residual gas scattering [9].

Figure 3(a) shows the probabilities from the model for various end products obtained using Gulley's approach. Predictions for current carbon foil thickness of 380 μgm/cm² and that proposed for PIP-II are also shown.

Figure 3(b) depicts the estimated probabilities for proton loss due to large angle Coulomb scattering during multi-turn injection in Booster with $BT$=10 and 14. For $BT$=14 in current scenario, we estimate ~ 0.3% beam loss arising from large angle Coulomb scattering indicating a considerable difference between experimental finding and our model estimate. This needs further investigation.

## SUMMARY AND FUTURE WORK

We revisited the beam loss due to foil scattering during injection for current Fermilab Booster operation and for future upgrades. The PMT-based fast-loss monitors have proved to be an invaluable tool in identifying the injection foil scattering loss. By correlating the PMT foil scattering signals to losses measured by the Booster toroid, our preliminary data analysis suggests ~1% beam losses at injection associated with foil-scattering.

A scattering model is presented as an attempt to understand beam the loss mechanism. We find that there is considerable discrepancy between the model and the measurements; the simple model underestimates the data by nearly a factor of three. We plan improve the model to better match the observations and, in the meantime, we are adding two more PMTs close to the foil to improve the measurements.